\documentclass[a4paper,twoside]{article}

\usepackage{epsfig}
\usepackage{subcaption}
\usepackage{calc}
\usepackage{amssymb}
\usepackage{amstext}
\usepackage{amsmath}
\usepackage{amsthm}
\usepackage{multicol}
\usepackage{pslatex}
\usepackage{apalike}
\usepackage[bottom]{footmisc}
\usepackage{SCITEPRESS}     

\begin{document}

\title{SAFR-AV: Safety Analysis of Autonomous Vehicles using Real World Data – An end-to-end solution for real world data driven scenario-based testing for pre-certification of AV stacks}

\author{\authorname{Sagar Pathrudkar\sup{1}, Saadhana B Venkataraman\sup{1}, Deepika Kanade\sup{1}, Aswin Ajayan\sup{1}, Palash Gupta\sup{1}, Shehzaman Salim Khatib\sup{1}, Vijaya Sarathi Indla\sup{1} and Saikat Mukherjee\sup{1}}
\affiliation{\sup{1}Siemens Technology, India}
\email{\{sagar.pathrudkar, saadhana.bvenkataraman\}@siemens.com}
}


\abstract{One of the major impediments in deployment of Autonomous Driving Systems (ADS) is their safety and reliability. The primary reason for the complexity of testing ADS is that it operates in an open world characterized by its non-deterministic, high-dimensional and non-stationary nature where the actions of other actors in the environment are uncontrollable from the ADS's perspective. This leads to a state space explosion problem and one way of mitigating this problem is by concretizing the scope for the system under test (SUT) by testing for a set of behavioral competencies which an ADS must demonstrate. A popular approach to testing ADS is scenario-based testing where the ADS is presented with driving scenarios from real world (and synthetically generated) data and expected to meet defined safety criteria while navigating through the scenario. We present SAFR-AV, an end-to-end ADS testing platform to enable scenario-based ADS testing. Our work addresses key real-world challenges of building an efficient large scale data ingestion pipeline and search capability to identify scenarios of interest from real world data, creating digital twins of the real-world scenarios to enable Software-in-the-Loop (SIL) testing in ADS simulators and, identifying key scenario parameter distributions to enable optimization of scenario coverage. These along with other modules of SAFR-AV would allow the platform to provide ADS pre-certifications.}

\onecolumn \maketitle \normalsize \setcounter{footnote}{0} \vfill

\section{\uppercase{Introduction}}
\label{sec:introduction}

Autonomous Vehicles promise many benefits over human drivers in terms of automation, safety, fuel economy and traffic efficiency. But to facilitate wide-spread acceptance and usage of autonomous vehicles, their safety and reliability needs to be proven. Some studies report 8.8 billion failure-free miles for the AV to be certified safe \cite{Kalra2016}. This is very expensive and unsafe to test in the real world. One way to mitigate this is to do pre-certifications using Simulation-in-the-loop (SIL) testing. The real world being a complex environment is difficult to model by hand. So, it becomes crucial to leverage data-driven methods and derive the AV environment models using real world data.
\cite{Kalra2016}.

\subsection{System Overview}
The behavioral competencies of autonomous vehicles capture their ability to perform appropriate driving behavior in various situations. But these competencies need to be tested thoroughly in all possible operating conditions that the AV would face in the real world. The Road-traffic ecosystem as well as the driving behaviors of other actors in an AV`s neighborhood create complex environments characterized by high-dimensionality (depends on large number of factors), nonlinearity (varies nonlinearly in a non-straightforward manner), high stochasticity (has random, unmodelled components) and non-stationarity (the underlying distributions are not constant and evolve over time). 
\begin{figure*}[!h]
  \centering
   \includegraphics[width = 12cm]{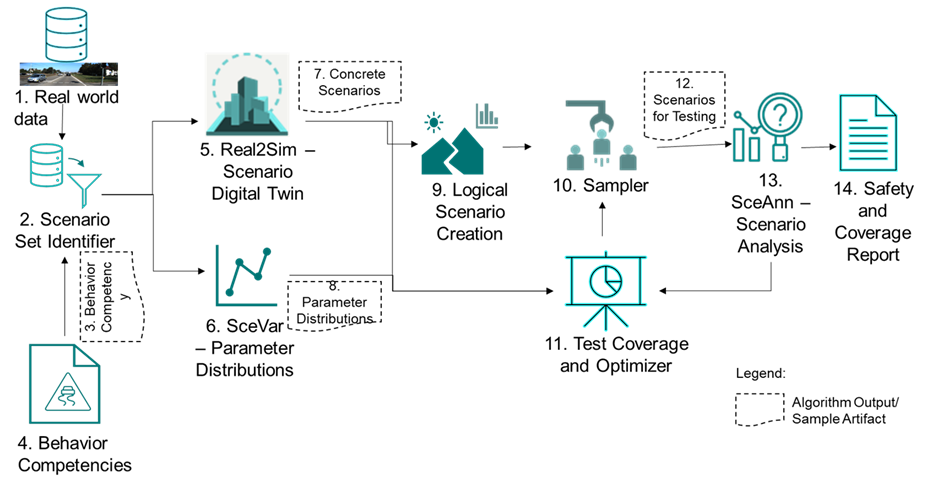}
  \caption{Workflow on Safety Analysis of Autonomous Vehicles from Real World Data or SAFR-AV}
  \label{fig:figure1}
\end{figure*}

This makes it very difficult to model the AV environment mathematically and computationally. The real-world data already accounts for the above complexities and thus can be used to model the AV environment. 
We present a solution called SAFR-AV (Safety Assessment from Real World Data for Autonomous Vehicles) (Figure~\ref{fig:figure1}) for end-to-end SIL testing of AV stacks wrt the behavior competencies.
The solution (Figure~\ref{fig:figure1}) involves several components: 

\begin{enumerate}
  \item Scenario Identifier (Block 2 in Figure~\ref{fig:figure1}) is used to identify and extract the scenarios relevant for a given behavior competency (Block 3) from real-world data (Block 1).
  \item The Real2Sim module (Block 5) converts real-world multi-sensor data to standard representations such as OpenScenario \cite{openscenario11}\cite{openscenario20} and creates a digital twin representation of the scenario (Block 7).
  \item The SceVar (Scenario Variations) module (Block 6) takes the real-world data of vehicle trajectories builds the probability distributions for different scenario parameters (Block 8). These distributions are used to construct logical scenarios (Block 9) which are then used to generate realistic synthetic variations of scenarios (Block 12) and compute coverage of test space (Block 11). These scenarios are then simulated in an AV simulator and the AV's response is then analyzed with respect to different automotive safety metrics. 
\end{enumerate}

\subsection{Key Contributions}
In this paper, we present three areas of work along with their corresponding results. 
\begin{enumerate}
\item Scenario Set Identifier: A large scale scenario set search engine to ingest 100's of Tb of data and search for specific behavioral competency scenarios
\item Real2Sim: A scenario digital twin creator to generate OpenScenario standard compliant concrete and smart scenarios from real world data	
\item SceVar: A statistical engine to generate parameter distributions for a given driving environment (also known as Operational Design Domain or ODD) which enables creation of logical scenarios and computation of coverage of parameter space
\end{enumerate}

\section{\uppercase{Prior Work}}
Scenario based testing is one of the standard approaches in AV testing \cite{Riedmaier2020}. The scenarios for AV testing typically have two sources: knowledge-driven \cite{Bagschik2018} that utilize domain knowledge from experts and standards and data-driven \cite{Fremont2020}\cite{MedranoBerumen2019} which utilize real-world data to get the ranges, distributions \cite{Knull2017}\cite{Wei2014}, and interactions of scenarios variables. 
Behavior Competency \cite{NHTSA2018} that demonstrate the ability of an AV to perceive, plan and act appropriately in a situation provide a way to concretize the testing requirements thus reducing the state space of AV scenarios. \cite{Tenbrock2021} presents a method for extracting scenario database by analyzing time series data from the perspective of every vehicle in the scene. Multi-stage perception pipelines have been proposed to identify objects and high-level events from video data \cite{Park2019}. Frameworks have been developed \cite{Karunakaran2022} to automatically build a dataset of logical lane change scenarios from sensor data, which can be used to sample test concrete scenarios.\\
\section{\uppercase{Scenario Set Identifier for Behavior Competencies}}
Behavioral Competency (BC) plays an important part in describing a common vocabulary for ADS competencies or features \cite{NHTSA2018}. The maturity of an ADS can be measured as the number of behavioral competencies that it can successfully and safely demonstrate. A key challenge in the scenario-based testing methodology is to identify a set of scenarios where the ADS can be tested for a given behavioral competency.
\begin{figure}[!h]
  \centering
   \includegraphics[width = 7cm]{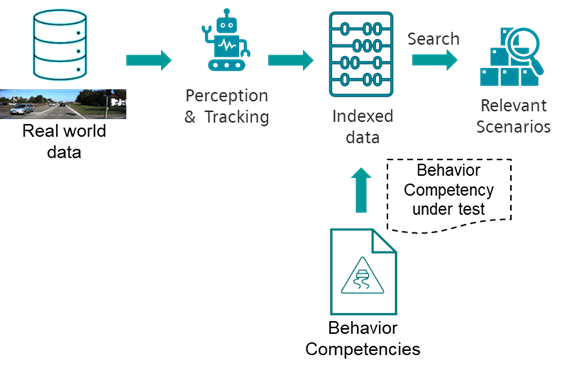}
  \caption{Scenario Set Identification for given behavioral competency from real world data}
  \label{fig:figure2}
\end{figure}
Our scenario set identifier tool allows users to search for such scenarios from data recorded in the real world and uploaded to the tool. Multi-sensor data (from cameras, lidar, radar and IMU sensors) is recorded and then perception algorithms (Object Detection, Tracking, Turn Detection, Intersection Identification) are run on the raw data to generate a novel metadata schema. The labelled metadata is indexed and stored in multiple SQL, document and time-series databases.
Behavior competencies are natural language descriptions of capabilities demonstrated in certain situations. We represent the behavioral competencies as a combination of map features (junctions, road sections, etc) and traffic events happening around the vehicle of interest (turns, lane changes, cut-in/ cut-out scenarios, etc). The behavioral competencies can be easily translated into search queries that employ the generated metadata to find the applicable scenarios. 
Sample datasets that have been tested include commonly available open-source AV datasets. Table \ref{tab:table1} below shows a subset of the behavioral competencies and their corresponding search queries to identify relevant scenarios where the ADS can be tested. 
\begin{table}[h]
\vspace{0.01cm}
\caption{Example search queries for behavior competencies}\label{tab:table1} \centering
\begin{tabular}{|p{2.5cm}|p{4cm}|}
  \hline
  \vspace{0.01cm}
  NHTSA Behavioral Competency (BC) & Example Search Query \\
  \hline
  \vspace{0.01cm}
  Detect and Respond to Lane Changes & $event=lane\_change$  \\
  \hline
  \vspace{0.01cm}
  Detect Traffic Signals and Stop/Yield Signs & $ODD.signage=stop$, $ODD.traffic\_signal=red$  \\
  \hline
  \vspace{0.01cm}
  Navigate Intersections and Perform Turns & $ODD.intersection = 3\-way$ \& $turn=left||right$  \\
  \hline
  \vspace{0.01cm}
  Perform High-Speed Merge (e.g., Freeway) & $ego_vehicle_event = merge$ \& $speed>50mph$ \& $ODD.road\_way\_type=freeway$  \\
  \hline
\end{tabular}
\end{table}

The key technical challenges that need to be addressed for building this search engine are a) high speed data ingestion pipeline to ingest Tb's of raw multi-sensor recorded data, b) algorithms for automatically labeling this data and c) indexing the annotated data for efficient sub-second latency search. 
We describe the key performance metrics for such a search engine – ingestion speed, search performance, and search accuracy. 
\begin{figure}[!h]
  \centering
   \includegraphics[width = 5cm]{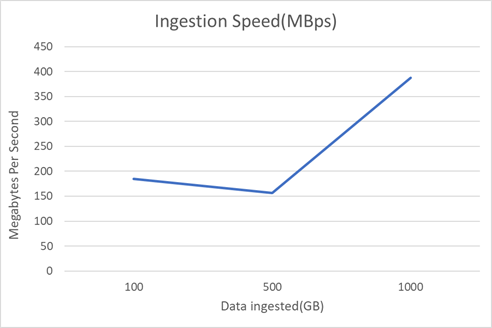}
  \caption{Ingestion speed vs Data ingested}
  \label{fig:figure3}
\end{figure}
\begin{figure}[!h]
  \centering
   \includegraphics[width = 5cm]{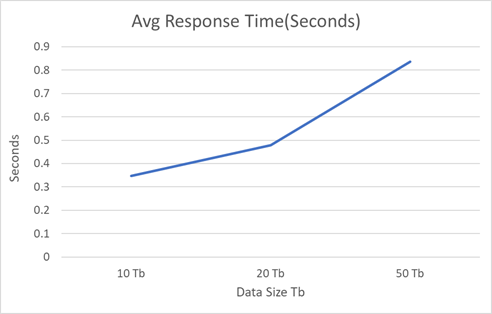}
  \caption{Response time vs Size of dataset searched}
  \label{fig:figure4}
\end{figure}
\begin{figure*}[!h]
  \centering
   \includegraphics[width = 12cm]{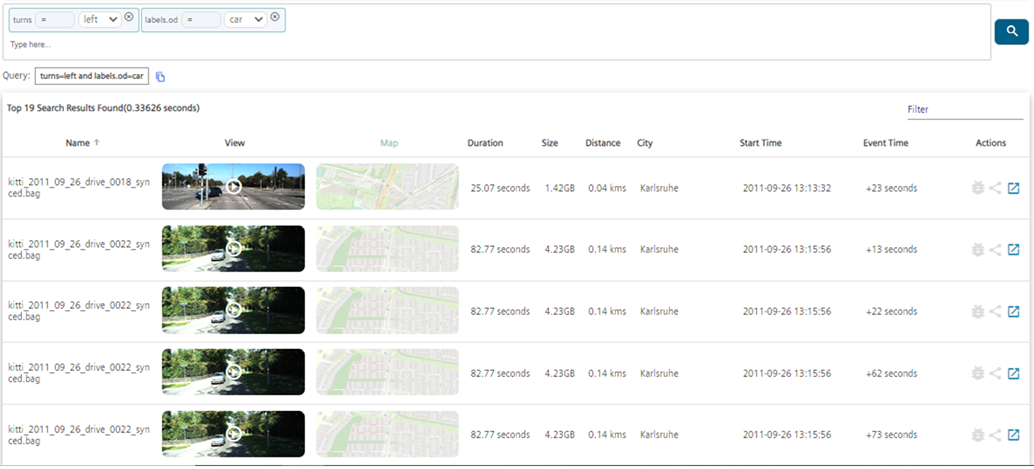}
  \caption{Example results for an AV Scenario Search query}
  \label{fig:figure5}
\end{figure*}
\begin{enumerate}
    \item Ingestion Speed: the ingestion speed describes the rate at which data can be ingested by the platform and is contextualized against the size of data ingested. For our tool (Figure~\ref{fig:figure2}) the Ingestion speed comes down slightly as we approach 500 GB of data ingested as we approach limits of vertical scaling, then it increases as we horizontally scale to more number of nodes. It can be seen in Figure~\ref{fig:figure3} that the ingestion speed improves for larger dataset size.
    \item Search Performance: The average search query response time describes how much time it'd take to search through a database and retrieve raw sensor data of scenarios relevant for the query. It can be seen in Figure~\ref{fig:figure4}, that our search performance is consistent, ie, the time taken to retrieve results from complexity analogous to O(n) where n is the size of database that is searched through. 
    \item Search Accuracy: When the behavioral competency is linked to a single search term (e.g. detect and respond to lane changes) the search accuracy is directly linked to the accuracy of the corresponding perception algorithm. However, for compound searches (e.g., perform high-speed merge) (Table \ref{tab:table1}) the accuracy of the search is computed as the product of the accuracy of independent search terms, as the events are detected independently.
\end{enumerate} 
Figure~\ref{fig:figure5} shows the scenario search tools which returns set of scenarios for a specific search query along with timestamps for when the queried event happened.
\section{\uppercase{Real2Sim - Real World to Simulation Digital Twin of Scenarios}}
There are three common ways to test for behavioral competencies \cite{Kalra2016} a) X in loop (X = Software, Hardware, Vehicle), b) testing in a closed-loop track and c) testing on public roads. In our approach we first use software in the loop testing (SIL) using simulators (e.g. CARLA, Siemens Prescan) to test the ADS stack.
\begin{figure}[!h]
  \centering
   \includegraphics[width = 7cm]{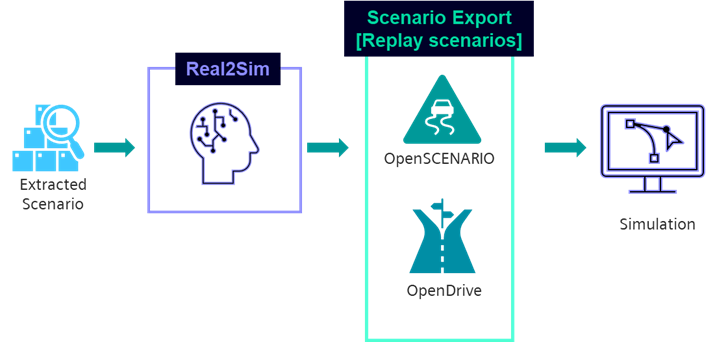}
  \caption{Digital Scenario Creation from Extracted Real World Data}
  \label{fig:figure6}
\end{figure}
The test scenarios for each behavioral competency identified from the module in Section III need to be converted to a digital twin for testing in a software simulator. We use a standard format (e.g. ASAM OpenScenario \cite{openscenario11}\cite{openscenario20}, OpenDrive \cite{opendrive17}) which is supported by a wide variety of simulators. This conversion from		 real world multi-sensor data to ASAM OpenSCENARIO is done using the Real2Sim module (Figure~\ref{fig:figure6}).
\subsection{Key Challenges in Real2Sim}
The key challenges in conversion from real world are fidelity and responsiveness. The fidelity of the conversion refers to “realism” of the digital twin wrt the static and responsive aspects of the environment. The challenge here is in detection and positioning the object relative to one another, especially if there are no high definition (HD) maps available. Fidelity also refers to the overall visual realism of the scenario, which calls for accurate positioning of street furniture, buildings and vegetation, along with realistic textures. 
Responsiveness refers to the elements such as vehicles, pedestrians, etc. that potentially react to their environment and stimuli that they receive. The the responsive elements of AV's environment should respond to the actions taken by the AV. This can be achieved by conditioning the actor behavior on ego actions through triggers. The OpenSCENARIO standard supports a set of triggers and conditions, and example of which is mentioned in Table \ref{tab:table2} below.
\begin{table}[h]
\caption{Trigger conditions that can be used to create responsive behavior}\label{tab:table2} \centering
\begin{tabular}{|p{2cm}|p{2cm}|p{2.2cm}|}
  \hline
  \vspace{0.01cm}
  Condition & Description & Example Application to Behavioral Competency \\
  \hline
  \vspace{0.01cm}
  Relative Distance Condition & Condition checking the relative distance between two entities & Other car to start taking a turn at the intersection when the ego vehicle is 50m from the intersection \\
  \hline
\end{tabular}
\end{table}
\subsection{Real2Sim Methodology}
The Real2Sim module automatically converts raw multi-sensor data into scenarios using the ASAM standard. There are the three important stages in the pipeline. 
\subsubsection{Multi-Sensor Situational Awareness}
A suite of algorithms are employed to extract the vehicle trajectories, road furniture, events and actions in the scene.The algorithms include 3D object detection for vehicles and road furniture (traffic cones, stop signs, etc), multi-object tracking, event and action detection (rapid deceleration, left turns, etc). We combine results from different redundant algorithms, based on different sensors like cameras, Lidar, GPS, IMU etc., to create a unified view of the situation in hand.
\begin{table}[h]
\caption{Elements of Scenario and evaluation metrics for quantifying the goodness of their digital twin}\label{tab:table3} \centering
\begin{tabular}{|p{1.7cm}|p{2.3cm}|p{2.2cm}|}
  \hline
  \vspace{0.01cm}
  Environment & Element & Accuracy Metric  \\
  \hline
  \vspace{0.01cm}
  Static & Buildings \& Traffic Furniture (e.g. lights, cones) & Intersection over Union (IoU) \\
  \hline
  \vspace{0.01cm}
  Static & Drivable surface area & Jacard Coefficient \\
  \hline
  \vspace{0.01cm}
  Static & Weather (wind, precipitation) & F1 Score \\
  \hline
  \vspace{0.01cm}
  Responsive & Non-player characters (cars, bicycles, trucks etc) & MOTA (Multiple Object Tracking Accuracy) \\
  \hline
  \vspace{0.01cm}
  Responsive & Traffic Lights & F1 score \\
  \hline
\end{tabular}
\end{table}
The accuracy of these algorithms is calculated using the corresponding standardized metrics, some of which are mentioned in Table \ref{tab:table3}.
\begin{table}[h]
\caption{Evaluation metrics for multi-sensor situational awareness algorithms on KITTI dataset \cite{kitti}}\label{tab:table4} \centering
\begin{tabular}{|p{1.7cm}|p{2.3cm}|p{2.2cm}|}
  \hline
  \vspace{0.01cm}
  Algorithm & Accuracy Metrics & Accuracy Numbers  \\
  \hline
  \vspace{0.01cm}
  3D Object Detection & Mean Average Precision (mAP) 
[mAP > 0.5]
 & 97.8\% [Car] \\
  \hline
  \vspace{0.01cm}
  Multiple Object Tracking & Multiple Object Tracking Accuracy (MOTA) & 65.17\% [Car] \\
  \hline
  \vspace{0.01cm}
  Detection of Turns & Precision and Recall
 & 1 and 0.6 \\
  \hline
  \vspace{0.01cm}
  Detection of Intersections & Precision and Recall & 1 and 1 \\
  \hline
\end{tabular}
\end{table}
The accuracy metrics for some of the algorithms used for multi-sensor situational awareness on the KITTI dataset is published in Table \ref{tab:table4}.
\subsubsection{Map Creation}
The static road structure of the objects are extracted in the ASAM OpenDRIVE format. While low fidelity versions are built with just open-source map information, the fidelity of the maps can be significantly improved by combining the map data with more 3D information.
\subsubsection{Scenario Export}
ASAM's OpenSCENARIO is being rapidly adopted as the standard for representing scenarios in the ADAS domain. Most tools and simulators in this domain support the format. Thus, representing the results from Situational Awareness (object lists, events, etc) in OpenSCENARIO's storyboard format is crucial in making the scenarios re-usable. The results from (a) are thus converted to stories, acts, triggers, conditions, maneuvers and maneuver groups and an OpenSCENARIO file is built as a result.
\subsection{Results}
We take a behavior competency under test: “detect intersections and take turns”. Using our Scenario Set Identification module we extract the scenes relevant to test this competency from real world data and use Real2Sim module to generate digital twin scenarios in standard format from these real-world scenes. (Figure~\ref{fig:figure5})
\begin{figure}[!h]
  \centering
   \includegraphics[width = 7cm]{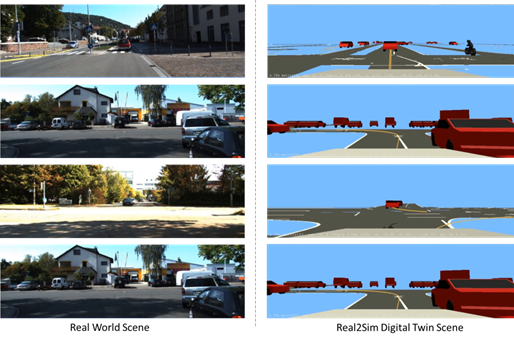}
  \caption{Digital Twins of the Scenarios from Real World Scenes relevant for a given Behavior competency (Detect Intersections and Take Turns) from KITTI Dataset \cite{kitti}}
  \label{fig:figure7}
\end{figure}
The Real2Sim module can be extended to support any dataset. Figure~\ref{fig:figure8} demonstrates a scenario created from PandaSet data. 
The scenario file from the Real2Sim module can be run by scenario engines from any supported tools or simulators. Figure~\ref{fig:figure9} demonstrates a Kitti scenario simulated in the Carla simulator.
\begin{figure}[!h]
  \centering
   \includegraphics[width = 7cm]{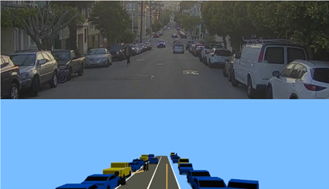}
  \caption{Real world scenario extracted from Pandaset \cite{pandaset} data and visualized in Esmini}
  \label{fig:figure8}
\end{figure}
\begin{figure}[!h]
  \centering
   \includegraphics[width = 7cm]{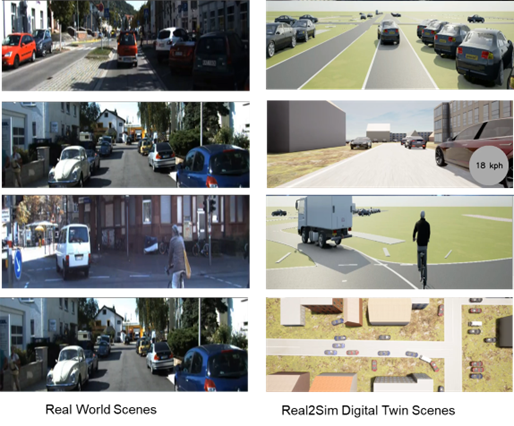}
  \caption{Scenario extracted from Kitti \cite{kitti} data simulated using Carla}
  \label{fig:figure9}
\end{figure}

\section{SceVar - Scenario Variations using Joint Parameter Distributions of Real-World Data}
\begin{figure}[!h]
  \centering
   \includegraphics[width = 7cm]{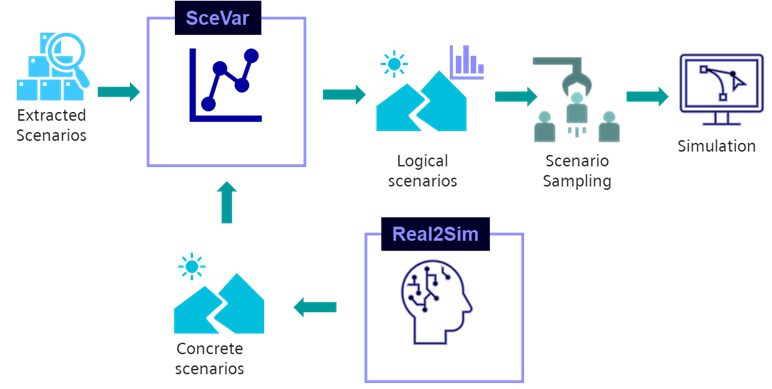}
  \caption{Identifying and Encoding distributions of real world behavior}
  \label{fig:figure10}
\end{figure}

We have discussed about methods to generate concrete scenarios or exact replays of the driving data recorded from real world in the Real2Sim section, where a scenario recorded from real world can be simulated and used for testing the AV stack. However there can be vast number of variations possible of a given scenario description and the AV should be exposed to them so as to cover the plausibilities it may encounter in the real world. Hence, we need to generate and sample all the plausible scenario variations (permutations and combinations of scenario variables) based on their statistical significance so as to maximize coverage of the test parameter(s) space. In terms of the ASAM standard \cite{openscenario11}, these statistical parameter distributions are used to create logical scenarios or scenarios which have a distribution instead of a fixed value. 
SceVar (contextualized in Figure~\ref{fig:figure10}) takes the object-list or trajectories (which involve timestamped coordinates of the vehicles and their velocities) data of all actors in the environment and constructs the joint probability distributions for the scenario parameters that together describe a scenario such as turning speed, turning angle, curvature, etc. used for the behavior competency under test. 
Constructing univariate distributions for the individual parameters would allow us to variate the values of that parameter and generate scenarios based on statistical significance of each value. But given that there are multiple scenario parameters that evolve together to create a scenario, one would need to generate joint probability distributions that describe how they behave together and thus generate even more realistic scenario variations.
\subsection{Results}
\begin{figure}[!h]
  \centering
   \includegraphics[width = 7cm]{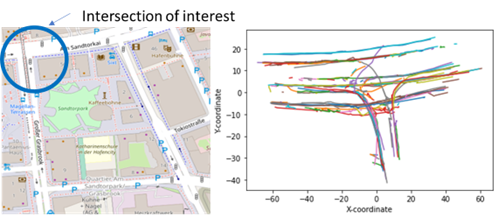}
  \caption{Map of geolocation and real world trajectories of vehicles driving}
  \label{fig:figure11}
\end{figure}

\begin{figure}[!h]
  \centering
   \includegraphics[width = 7cm]{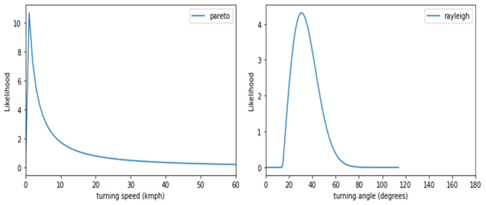}
  \caption{Univariate parameter distributions for turning speed and turning angles}
  \label{fig:figure12}
\end{figure}

\begin{figure}[!h]
  \centering
   \includegraphics[width = 4cm]{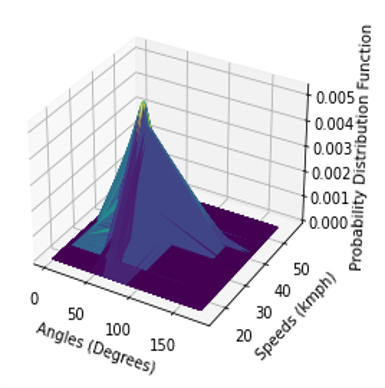}
  \caption{Joint probability distributions of turning speed and turning radius}
  \label{fig:figure13}
\end{figure}

\begin{figure}[!h]
  \centering
   \includegraphics[width = 5cm]{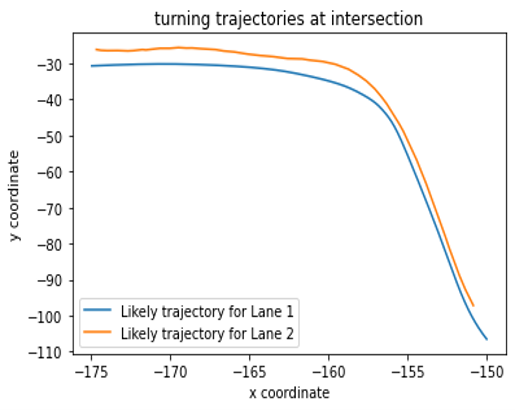}
  \caption{Statistically normal turning trajectories learned from real world data}
  \label{fig:figure14}
\end{figure}

\begin{figure}[!h]
  \centering
   \includegraphics[width = 6cm]{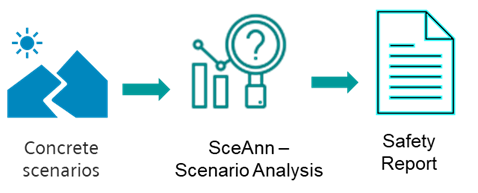}
  \caption{Analyzing Scenarios and AV's safety response}
  \label{fig:figure15}
\end{figure}

\begin{figure}[!h]
  \centering
   \includegraphics[width = 6cm]{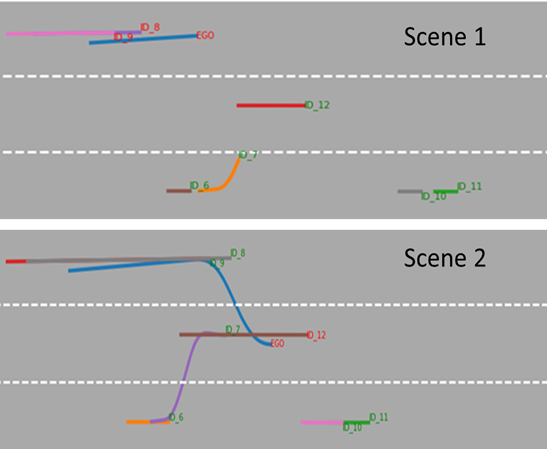}
  \caption{Per-scene Classification of each AV-other vehicle interaction as safe or unsafe}
  \label{fig:figure16}
\end{figure}

Continuing from the behavior competency “Detect Intersections and Take Turns”, the parameters of interest would be turning speed, turning angle (Figure~\ref{fig:figure12}, Figure~\ref{fig:figure13}) and turning trajectory (Figure~\ref{fig:figure14}). These are learned using statistical methods on large amount of real-world data. The SceVar module requires map information (e.g., OpenDrive \cite{opendrive17}) along with the timestamped coordinates of the vehicles in that geolocation as shown in Figure~\ref{fig:figure11}.
Once the scenario variations are simulated with the AV stack in the loop in an AV environment simulator such as Siemens Prescan or CARLA, the resulting simulation data is analyzed with respect to different safety metrics such as Time-to-collision (TTC) to quantify the safety performance of the AV stack using the Scenario Analyzer (SceAnn) module (Figure~\ref{fig:figure15}). 
Figure~\ref{fig:figure16} shows snapshots of two scenes where each vehicle's interaction is termed safe (vehicle name is in red color font) and unsafe (vehicle name is in green colored font). In the first scene, the ego vehicle's interaction with vehicles ID\_8 and ID\_9 is unsafe. The ego vehicle then changes lanes and moves to another lane and the scene becomes safe with respect to these two vehicles, but becomes unsafe with respect to the ID\_12 vehicle. These scores from each scene are then aggregated at a scenario level to generate assessment on the AV's safety performance in that scenario.
\section{Future Work and Research Challenges}
In this work, we presented SAFR-AV, a platform for end-to-end Simulation-in-the-loop (SIL) testing of AV stacks using real-world data for the purpose of pre-certification wrt various behavior competencies. We presented results on extraction of real-world scenes relevant for a behavioral competency under test, conversion of these scenes into digital twin representations and generation of real-world distribution of scenario parameters for optimizing test coverage. 
Future work would involve development of coverage optimization and smart sampling engines for ensuring exposure of statistical variability of the test scenario and to generate edge/critical cases. 
The research challenges include: 
\begin{enumerate}
    \item Robustness and accuracy of the multi-sensor perception algorithms used to extract relevant scenarios from real world data. 
    \item Robust map-matching to position the objects correctly. 
    \item Generating multi-variate heterogeneous probability distributions of sets of scenario parameters. 
    \item Causality Analysis in the safety assessment module to identify failure points and modes for the AV stack.
\end{enumerate}

\nocite{*}
\bibliographystyle{apalike}
{\small
\bibliography{main}}

\begin{thebibliography}{}

\bibitem[{ASAM}, 2017]{opendrive17}
{ASAM} (2017).
\newblock {ASAM OpenDrive 1.7.0 User Guide}.
\newblock ASAM Standard ASAM OpenDRIVE 1.7.0, {Association for Standardization
  of Automation and Measuring Systems (ASAM)}.
\newblock [Online; accessed 15-February-2023].

\bibitem[{ASAM}, 2018]{openscenario11}
{ASAM} (2018).
\newblock {ASAM OpenSCENARIO 1.1 User guide}.
\newblock ASAM Standard ASAM SCENARIO 1.1, {Association for Standardization of
  Automation and Measuring Systems (ASAM)}.
\newblock [Online; accessed 15-February-2023].

\bibitem[{ASAM}, 2022]{openscenario20}
{ASAM} (2022).
\newblock {ASAM OpenSCENARIO 2.0 Public release candidate}.
\newblock ASAM Standard ASAM SCENARIO 2.0, {Association for Standardization of
  Automation and Measuring Systems (ASAM)}.
\newblock [Online; accessed 15-February-2023].

\bibitem[Bagschik et~al., 2018]{Bagschik2018}
Bagschik, G., Menzel, T., and Maurer, M. (2018).
\newblock Ontology based scene creation for the development of automated
  vehicles.
\newblock In {\em 2018 IEEE Intelligent Vehicles Symposium (IV)}, pages
  1813--1820.

\bibitem[Clarke et~al., 2011]{Clarke2011}
Clarke, E.~M., Klieber, W., Nováček, M., and Zuliani, P. (2011).
\newblock Model checking and the state explosion problem.
\newblock In {\em LASER Summer School on Software Engineering}, pages 1--30,
  Berlin, Heidelberg. Springer.

\bibitem[Fremont et~al., 2020]{Fremont2020}
Fremont, D.~J., Kim, E., Pant, Y.~V., Seshia, S.~A., Acharya, A., Bruso, X.,
  and Mehta, S. (2020).
\newblock Formal scenario-based testing of autonomous vehicles: From simulation
  to the real world.
\newblock In {\em 2020 IEEE 23rd International Conference on Intelligent
  Transportation Systems (ITSC)}, pages 1--8.

\bibitem[Geiger et~al., 2013]{kitti}
Geiger, A., Lenz, P., Stiller, C., and Urtasun, R. (2013).
\newblock {Vision meets Robotics: The KITTI Dataset}.
\newblock In {\em International Journal of Robotics Research}, volume~32, pages
  1231--1237.
\newblock [Online; accessed 15-February-2023].

\bibitem[Kalra and Paddock, 2016]{Kalra2016}
Kalra, N. and Paddock, S.~M. (2016).
\newblock Driving to safety: How many miles of driving would it take to
  demonstrate autonomous vehicle reliability?

\bibitem[Karunakaran et~al., 2022]{Karunakaran2022}
Karunakaran, D., Berrio, J.~S., Worrall, S., and Nebot, E. (2022).
\newblock Automatic lane change scenario extraction and generation of scenarios
  in openx format from real-world data.
\newblock {\em Robotics and Autonomous Systems}, 148:103816.

\bibitem[Knull, 2017]{Knull2017}
Knull, J.~E. (2017).
\newblock Turn detection and analysis of turn parameters for driver
  characterization.

\bibitem[Medrano-Berumen and Akbaş, 2019]{MedranoBerumen2019}
Medrano-Berumen, C. and Akbaş, M.~I. (2019).
\newblock Abstract simulation scenario generation for autonomous vehicle
  verification.
\newblock In {\em 2019 SoutheastCon}, pages 1--6.

\bibitem[{National Highway Traffic Safety Administration (NHTSA)},
  2018]{NHTSA2018}
{National Highway Traffic Safety Administration (NHTSA)} (2018).
\newblock A framework for automated driving system testable cases and
  scenarios.

\bibitem[Park et~al., 2019]{Park2019}
Park, J., Wen, M., Sung, Y., and Cho, K. (2019).
\newblock Multiple event-based simulation scenario generation approach for
  autonomous vehicle smart sensors and devices.
\newblock In {\em 2019 IEEE Intelligent Transportation Systems Conference
  (ITSC)}, pages 1003--1008.

\bibitem[Riedmaier et~al., 2020]{Riedmaier2020}
Riedmaier, S., Ponn, T., Ludwig, D., Schick, B., and Diermeyer, F. (2020).
\newblock Survey on scenario-based safety assessment of automated vehicles.
\newblock {\em IEEE Access}, 8:87456--87477.

\bibitem[Tenbrock et~al., 2021]{Tenbrock2021}
Tenbrock, A., König, A., Keutgens, T., Bock, J., Weber, H., Krajewski, R., and
  Zlocki, A. (2021).
\newblock The conscend dataset: Concrete scenarios from the highd dataset
  according to alks regulation unece r157 in openx.

\bibitem[Wei et~al., 2014]{Wei2014}
Wei, F., Guo, W., Liu, X., Liang, C., and Feng, T. (2014).
\newblock Left-turning vehicle trajectory modeling and guide line setting at
  the intersection.
\newblock {\em Discrete Dynamics in Nature and Society}.

\bibitem[Xiao et~al., 2021]{pandaset}
Xiao, P., Shao, Z., Hao, S., Zhang, Z., Chai, X., Jiao, J., Li, Z., Wu, J.,
  Sun, K., Jiang, K., Wang, Y., and Yang, D. (2021).
\newblock {PandaSet: Advanced Sensor Suite Dataset for Autonomous Driving}.
\newblock {\em arXiv preprint arXiv:2112.06058}.
\newblock [Online; accessed 15-February-2023].

\end{thebibliography}

\end{document}